\documentclass[12pt,preprint]{aastex}

\newcommand{\nai}{\hbox{Na}~{\sc i}}
\newcommand{\mgii}{{\rm Mg}~{\sc ii}}
\newcommand{\caii}{\hbox{Ca}~{\sc ii}}
\newcommand{\hi}{\hbox{H}~{\sc i}}
\newcommand{\mgi}{\hbox{Mg}~{\sc i}}

\begin{document}


\title{First Detection of \nai\ D lines in High-Redshift Damped
Lyman-$\alpha$ Systems \footnote{Based on data collected at Subaru Telescope,
which is operated by the National Astronomical Observatory of
Japan.}}

\author{
SOHEI KONDO\altaffilmark{2},
NAOTO KOBAYASHI\altaffilmark{2},
YOSUKE MINOWA\altaffilmark{2},
TAKUJI TSUJIMOTO\altaffilmark{3},
CHRISTOPHER W. CHURCHILL\altaffilmark{4},
NARUHISA TAKATO\altaffilmark{5}, 
MASANORI IYE\altaffilmark{3}, 
YUKIKO KAMATA\altaffilmark{3},
HIROSHI TERADA\altaffilmark{5}, 
TAE-SOO PYO\altaffilmark{5}, 
HIDEKI TAKAMI\altaffilmark{5}, 
YUTAKA HAYANO\altaffilmark{5}, 
TOMIO KANZAWA\altaffilmark{5},
D. SAINT-JACQUES\altaffilmark{6},
WOLFGANG G\"{A}ESSLER\altaffilmark{7}, 
SHIN OYA\altaffilmark{5}, 
KO NEDACHI\altaffilmark{2}, 
ALAN TOKUNAGA\altaffilmark{8}
}

\altaffiltext{2}{Institute of Astronomy, University of Tokyo, 2-21-1
Osawa, Mitaka, Tokyo, 181-0015, Japan;\ kondo@ioa.s.u-tokyo.ac.jp}

\altaffiltext{3}{National Astronomical Observatory of Japan, 2-21-1
Osawa, Mitaka, Tokyo 181-8588, Japan.}

\altaffiltext{4}{Department of Astronomy, New Mexico State University,
1320 Frenger Mall, Las Cruces, NM 88003-8001}

\altaffiltext{5}{Subaru Telescope, National Astronomical Observatory
of Japan, 650 North A'ohoku Place, Hilo, HI 96720.}

\altaffiltext{6}{Groupe d'astrophysique, Universit\'{e} de
Montr\'{e}al, 2900 Boulevard \'{E}douard-Montpetit, Montr\'{e}al, QC
H3T 1J4, Canada}

\altaffiltext{7}{Max-Plank Institute f\"{u}r Astronomie,
K\"{o}nigstuhl 17, 69117 Heidelberg, Germany }

\altaffiltext{8}{Institute for Astronomy, University of Hawaii, 2680
Woodlawn Drive, Honolulu, HI 96822}

\begin{abstract}

A Near-infrared $\rm{(1.18-1.35\ \mu m)}$ high-resolution spectrum of
the gravitationally-lensed QSO APM 08279+5255 was obtained with the
InfraRed Camera and Spectrograph mounted on the Subaru Telescope using
the adaptive optics system. We detected strong \nai\ D $\rm{\lambda
\lambda\ 5891, 5897}$ doublet absorption in high-redshift damped
Ly$\alpha$ systems (DLAs) at $\rm{z_{abs}=1.062\ and\ 1.181}$,
confirming the presence of \nai, which was first reported for the
rest-frame UV \nai\ $ \rm{\lambda\lambda\ 3303.3, 3303.9}$ doublet by
Petitjean et~al. This is the first detection of \nai\ D absorption in a
high-redshift $\rm{(z>1)}$ DLA.  In addition, we detected a new \nai\
component in the $\rm{z_{abs}=1.062}$ DLA and four new components in the
$\rm{z_{abs}=1.181}$ DLA.  Using an empirical relationship between \nai\
and \hi\ column density, we found that all components have large \hi\
column density $\rm{(\log\, N_{HI}\ [cm^{-2}] \gtrsim 20.3 )}$, so that
each {\it component\/} is classified as DLA absorption.
We also detected strong \nai\ D absorption associated with a \mgii\
system at $\rm{z_{abs}=1.173}$.  Because no other metal absorption lines
were detected in this system at the velocity of the \nai\ absorption in
previously reported optical spectra (observed 3.6 years ago), we
interpret this \nai\ absorption cloud probably appeared in the line of
sight toward the QSO after the optical observation.  This newly found
cloud is likely to be a DLA based upon its large estimated \hi\ column
density.  We found that the $\rm{N_{NaI}/N_{CaII}}$ ratios in these DLAs
are systematically smaller than those observed in the Galaxy; they are
more consistent with the ratios seen in the Large Magellanic Cloud.
This is consistent with dust depletion generally being smaller in lower
metallicity environments.  However, all five clouds of the
$\rm{z_{abs}=1.181}$ system have a high $\rm{N_{NaI}/N_{CaII}}$ ratio,
which is characteristic of cold dense gas.  We tentatively suggest that
the host galaxy of this system may be the most significant contributor
to the gravitational-lens toward APM 08279+5255.

\end{abstract}


\section{Introduction}

Damped Ly$\alpha$ systems (DLAs) comprise a class of the QSO absorption
line systems characterized by high \hi\ column density greater than $2\times
10^{20}\ \rm{[cm^{-2}]}$ \citep{wol86}.  DLAs are one of the
strongest probes to examine the evolution of metallicity and dust
depletion in gaseous components of high redshift galaxies
\citep[e.g.,][]{Pet04}.
Since there is a well-established correlation between $N_{\rm{Na\ I}}$ and
$N_{\rm{HI}}$ \citep{Fer85,Bow95}, \nai \ absorption, which has been
observed extensively in the Galaxy, holds the potential to estimate
$N_{\rm{H\ I}}$ for individual clouds in DLAs, which usually cannot be
precisely measured because of heavy blending of \hi\ absorption.
Moreover, the column density ratio of \nai\ to \caii\ provides a useful  
indicator of the degree of dust depletion of ISM clouds
\citep[e.g.,][]{Wels90,Cra92,Ber93,Sem94,Pri01,Cra02} because sodium
is hardly depleted from gas onto dust, while calcium can be heavily
depleted \citep{Sav96}.

The main absorption lines of \nai\ are the optical $\lambda\lambda\
5891, 5897$ D doublet and the ultraviolet $\lambda\lambda\ 3303.3,
3303.9$ doublet.  Observations exploiting the \nai\ D doublet hold the
greater promise for comprehensive studies using \nai\ due to two
important advantages: (1) the oscillator strength of the \nai\ D
transitions are $\sim70$ times greater than the \nai\ UV transitions
\citep{Mor03}, and (2) \nai\ D absorption rarely overlaps with other
metal absorption lines because the wavelengths of the \nai\ D doublet
are significantly redder than most of other metal lines.  Thus, \nai\ UV
doublets can be detected only in the highest column density systems
toward the brightest QSO.  Weak \nai\ absorption systems are more
likely to be detected with the \nai\ D doublet.  However, observations
of \nai\ D in high-redshift DLAs require near-infrared high resolution
spectroscopy because the rest-frame optical wavelengths are redshifted
into the near-infrared. The advent of high sensitivity and high
resolution spectroscopy in the near-infrared with 8-meter class
telescopes has enabled us to undertake such observations
\citep[e.g.,][]{Kob02,Kob03,Kob05}.

As a first trial for detecting \nai\ D absorption lines at high
redshift ($z > 1$), we observed the extremely bright
$\rm{z_{em}=3.911}$ gravitationally-lensed QSO "APM 08279+5255"
\citep{Irw98}.  There are two dominant gravitationally-lensed images,
"A" and "B," with a separation of $0\farcs 38$ and flux ratio of
$f_B/f_A = 0.77$ \citep{Iba99}. A third faint image, "C", was also
reported by Ibata et~al.\ \nocite{Iba99} between A and B, with a
separation of A - C = $0\farcs 15$ and flux ratio $f_C/f_A =
0.18$. \citet{Pet00} reported probable faint \nai\ UV doublet
absorption associated with the $\rm{z_{abs}}=1.062$ and 1.181 DLAs in
optical HIRES \citep{Vogt94} spectra obtained by
\citet{Ell99a,Ell99b}.  For both systems, Petitjean et~al.\ concluded
(1) $\log\, N_{\rm{HI}} \sim21$, (2) metallicity $0.3-1.0\,
Z_{\odot}$, (3) dust-to-metal ratio about half that of the Galaxy, and
(4) a temperature of $\sim$few hundred Kelvin.

In this paper, we present near-infrared 1.18 -1.35 $\rm{\mu}$m spectra
of APM 08279+5255 with a resolution of about 50 $\rm{km\ s^{-1}}$.
The spectra were obtained with the InfraRed Camera and Spectrograph
(IRCS) using the adaptive optics (AO) system at the Japanese Subaru
Telescope.  The AO improved the FWHM of both images A
and B such that the spectra of A and B were marginally separated. However, we combined the
spectra of images A and B in order to maximize the signal-to-noise
ratio and to compare our results with those by
\citet{Pet00}. We will publish A-B separated spectra elsewhere.
This paper is structured as follows: In \S 2, we describe the IRCS
observations and data reduction of APM 08279+5255. In \S 3, we present
the \nai\ D absorption lines and the fitting results for the
$\rm{z_{abs}=1.062,1.173,1.181}$ systems.  We discuss our results and
conclusions in \S 4.

\section{Observation \& Data Reduction}

Near-infrared 1.18-1.35 $\rm{\mu m}$ spectra of APM
08279+5255 were obtained on 2001 December 25 UT, during the scientific
commissioning run of the IRCS instrument.  IRCS is a cassegrain
mounted 0.9-5.5 $\rm{\mu m}$ near-infrared camera and spectrograph
\citep{Tok98,Kob00} with an AO system \citep{Tak04} on the 8.2m  
Subaru Telescope \citep{iye04} located on Mauna Kea, Hawaii.

The Subaru AO system uses a curvature wave front sensor with 36
control elements to compensate for a disturbed wavefront due to the
earth's atmosphere. For an $R \sim 11$ wavefront reference star under
excellent observing conditions, the AO provides a stellar image with a
Strehl ratio of $\sim 0.04$ and a FWHM of $\sim 0\farcs 07$ in the
$J$-band $\rm{(1.25\ \mu m)}$.  The performance degrades with
fainter reference stars.  For these observations, we used APM
08279+5255 itself ($R \sim 15.6$) as the wavefront reference star.  We
achieved a FWHM of 0\farcs2 in the $J$-band, which is much narrower
than (0\farcs5$-$0\farcs6) for typical seeing conditions at the Subaru
Telescope.

We used the cross-dispersed Echelle mode of the IRCS, which provides a
spectral resolution of 15 km $\rm{s^{-1}}$ $(\lambda/\Delta \lambda\sim
20000)$ with a 0\farcs15 slit. The pixel scale is 0\farcs075
$\rm{(\sim7.5\ km\ s^{-1})}$ per pixel across the slit and 0\farcs060
per pixel along the slilt. The entire $J$ -band $\rm{(1.18-1.35\ \mu
m)}$ was covered simultaneously in one IRCS exposure, with the
cross-disperser in order 5 and the echelle in orders 42-48.  Although
the FWHM of the images were good throughout the observing time
(about 0\farcs 2-0\farcs5 in the $J$-band), we used a
wider slit (0\farcs 6) to maximize throughput. Thus, the
spectral resolution was determined by the average image size $\sim$
0\farcs 5, and the resulting spectral resolution was $\rm{\sim 50\ km\
s^{-1}\ (\lambda /\Delta \lambda \sim 6000)}$ in the observed wavelength
range.

The slit length was $3\farcs 8$, and the position angle was set to
32\degr\ so that both the lensed images A and B are in the slit.  The
telescope pointing  was nodded along the slit by about 2\arcsec\ for each 600s
exposure for sky-background and dark subtraction.  Because of the fine
pixel scale needed for AO images and the high spectral resolution, it
is impossible to reach the background limit in 600s exposures, which
are required to be less than the time variation of the sky OH
emission. Eight sets of data were obtained, resulting in a total
exposure time of 9600s.  Spectra of bright telluric standard stars (HD
92728: A0Vs) at similar air mass were obtained in a similar fashion.

All the data were reduced following standard procedures using the
IRAF\footnote{IRAF is distributed by the National Optical Astronomy
Observatories, which are operated by the Association of Universities for
Research in Astronomy, Inc., under cooperative agreement with the
National Science Foundation.}  {\it noao.imred.echelle} package,
including sky subtraction (subtraction of two frames), flat-fielding
(halogen lamp with an integrating sphere), and aperture extraction.
Argon lamp spectra that were taken at the end of the observing night
were used for vacuum wavelength calibration. A
Heliocentric correction of $\rm{9.64\ km\ s^{-1}}$ was applied to the
vacuum wavelength calibrated spectra.

\section{Results}
\begin{figure}
\plottwo{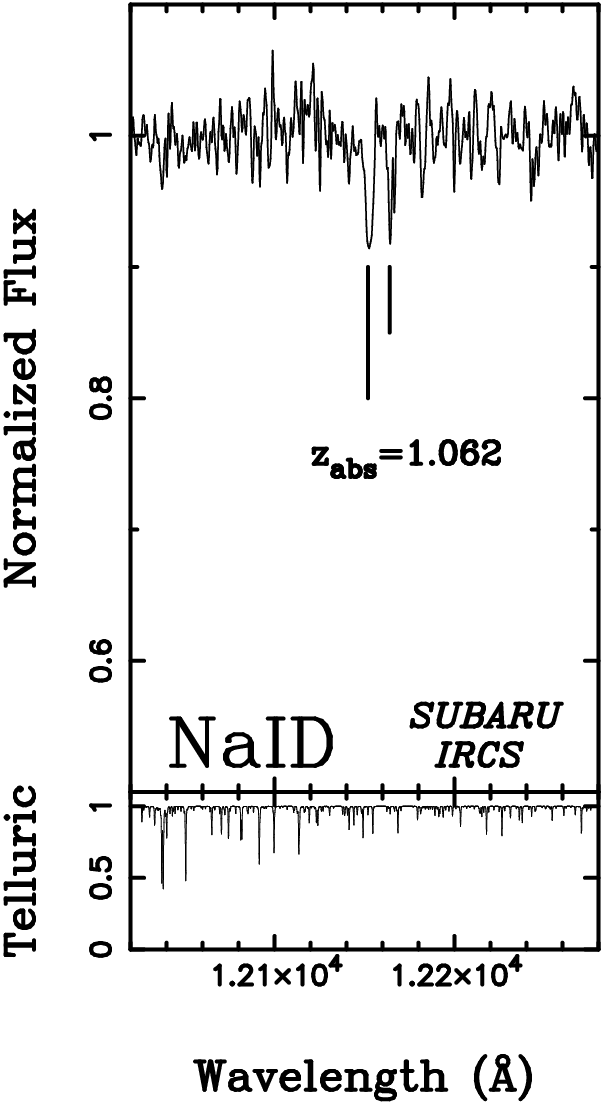}{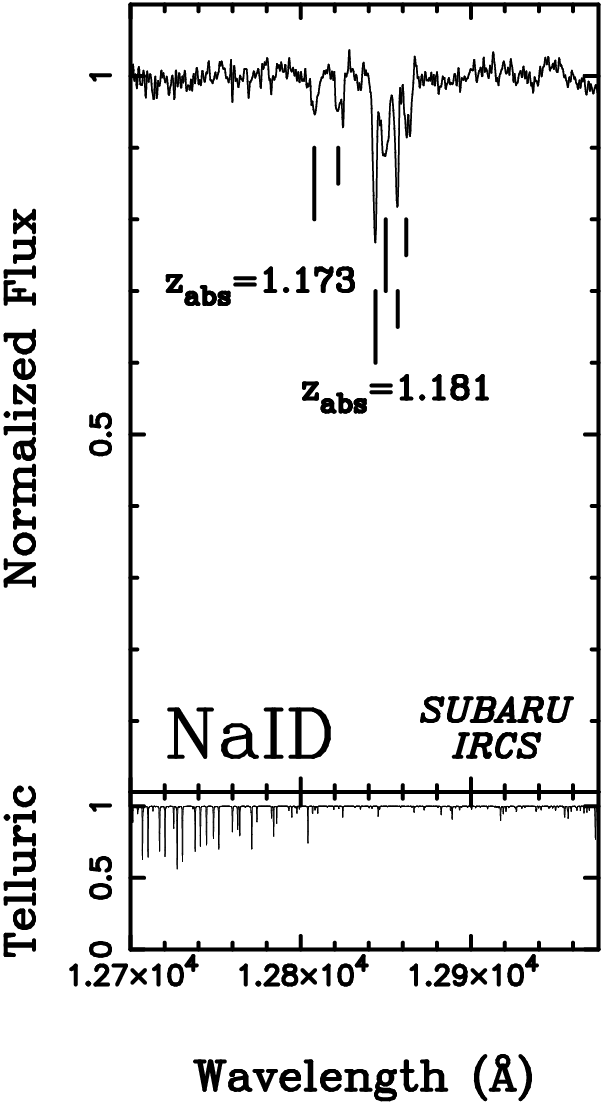}
 \caption{Detected \nai\ D absorption lines at $\rm{z_{abs}=1.062}$
 (left panel) and $\rm{z_{abs}=1.173, 1.181}$ (right panel). Vertical
 axis is normalized flux to the continuum level and horizontal axis is
 local-frame wavelength in both panels.  The spectrum was smoothed with
 a 3 pixel $ \rm{(\sim 20\ km\ s^{-1})}$ boxcar function. Longer lines
 indicate \nai\ D $\lambda5891$, while shorter lines indicate \nai\ D
 $\lambda5897$.  The left spectrum is noisier than the right one because
 the \nai\ D absorption lines at $\rm{{z_{abs}=1.062}}$ were located on
 the edge of the detector where the detector noise is significantly
 larger.  The bottom panel shows the telluric absorption spectrum
 calculated with ATRAN package software \citep{Lor92}.\label{fig1}}
\end{figure}

In Figure \ref{fig1}, we present the $\rm{1.18-1.35\ \micron}$ spectrum
of APM 08279+5255. The raw spectrum was smoothed with a 3 pixel
$(\rm{\sim 20\ km\ s^{-1}})$ boxcar function for
presentation. \nai\ D absorption lines from the
$\rm{z_{abs}}=1.062$ and 1.181 DLAs are clearly seen. Several velocity
components can be seen in the \nai\ D absorption lines from the
$\rm{z_{abs}=1.181}$ DLA. Additionally \nai\ D absorption lines
associated with the $\rm{z_{abs}=1.173}$\ \mgii\ system \citep{Pet00}
are detected. This doublet absorption line was identified as \nai\ D
lines because other metal lines of other absorption systems listed in
Ellison et al. (1999a,b) cannot account for these two absorption
features.

We fit Voigt profiles to the \nai\ D absorption features assuming the
component velocities identified by \citet{Pet00} from metal absorption
lines in the optical spectra by \citet{Ell99a,Ell99b}.  The column
density, Doppler width, and redshift of each component were evaluated
using VPGUESS\footnote{VPGUESS is a graphical interface to VPFIT written
by Jochen Liske, \url{http://www.eso.org/$\tilde{\ }$jliske/vpguess/}}
and VPFIT\footnote{VPFIT is a Voigt profile fitting package provided by
Robert F. Carswell, \url{http://www.ast.cam.ac.uk/$\tilde{\
}$rfc/vpfit.html}} \citep{Car87}. We fit two components for the
$\rm{z_{abs}=1.062}$ system and five for the $\rm{z_{abs}=1.181}$
system. The absorption profiles are dominated by the
instrument profile, which is almost identical to the Gaussian width
FWHM$\ \sim 50\ \rm{km\ s^{-1}}$.

For the $\rm{z_{abs} = 1.173}$ system, there is no a
priori information on velocity components unlike the other
systems. Also, there are several adjacent hot pixels on the right side
of the \nai\ 5897 absorption line, which makes the fitting with multi
velocity components difficult. Those hot pixels were difficult to remove
because of their proximity to the  \nai\ absorption line (see the feature
labeled with $\times$ mark in Figure \ref{fig4}). Because of the above
reasons, we performed only one component VPFIT (almost equivalent to a
simple Gaussian fitting) for this system. The hot pixel regions were
ignored for the fitting. In view of the statistical noise that can be
estimated by point-to-point fluctuations on the continuum, the single
component fitting is almost satisfactory.

\begin{figure}
\epsscale{.60} \plotone{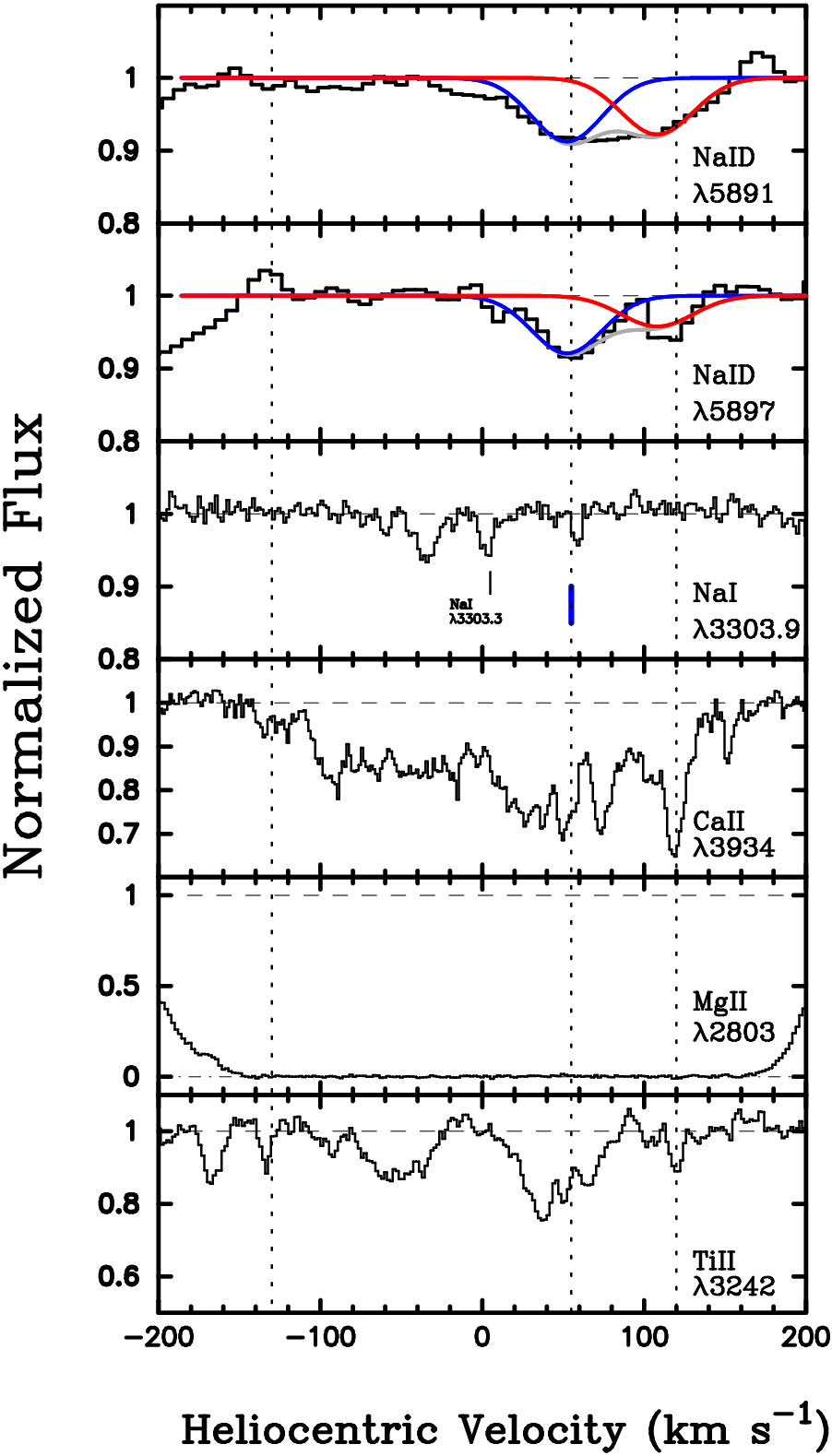} \caption{Velocity profile of metal lines
in $\rm{z_{abs}=1.062}$ system. Horizontal axis shows the heliocentric
velocity from z=1.06230. Vertical axis is normalized flux to the
continuum level. Top two panels show the velocity profiles of \nai\ D
$\lambda\lambda5891, 5897$ absorption lines obtained with IRCS with the
velocity resolution of $\sim$ 50 km $\rm{s^{-1}}$. Bottom four panels
show the velocity profile of \nai\ $\lambda 3303.9$, \caii\ $\lambda
3934$, \mgii, $\lambda 2803$ and {\rm Ti}~{\sc ii}\ $\lambda 3242$
absorption lines obtained with HIRES with the velocity resolution of
$\sim$ 8 km $\rm{s^{-1}}$. Vertical dashed lines mark the velocity
components found by \citet{Pet00}. The blue line shows
the \nai\ velocity component which has been detected with \nai\ $\lambda
\lambda 3303.3, 3303.9$ \citep{Pet00} and the red line shows
the newly detected \nai\ velocity component determined from the VPFIT
program, which was also seen in the \caii, \mgii, and {\rm Ti}~{\sc ii}
line profiles. Gray lines show the sum of all components. In the HIRES
spectra, unmarked absorption lines are either from telluric absorption
or other unrelated systems.\label{fig2}}
\end{figure}

\begin{figure}
\epsscale{.60}
\plotone{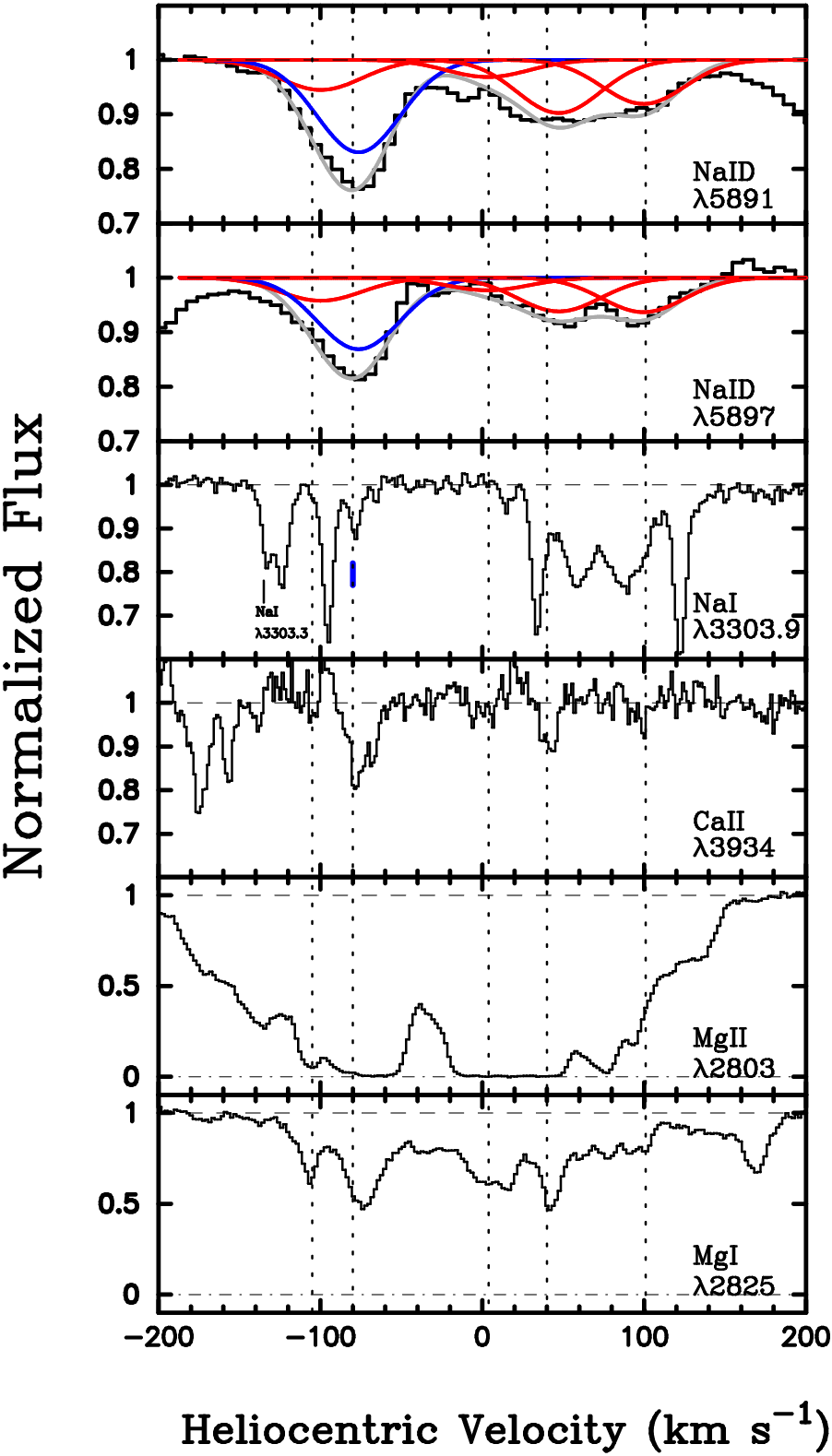}
 \caption{Velocity profile of metal lines in the $\rm{z_{abs}=1.18}$
 system. Notations are same as those in Figure \ref{fig2}. The horizontal axis
 shows the heliocentric velocity from z=1.1807.  Bottom four panels show
 the velocity profile of \nai\ $\lambda 3303.9$, \caii\ $\lambda 3934$,
 \mgii\ $\lambda 2803$ and {\rm Mg}~{\sc i} $\lambda 2825$ absorption
 lines obtained with HIRES. Vertical dashed lines mark velocity
 components $\rm{(V=-105,-80,+40\ km\ s^{-1})}$ found by \citet{Pet00}
 and those $\rm{(V=+4,+101\ km\ s^{-1})}$ newly found with \nai\ D data.\label{fig3}}
\end{figure}

\begin{figure}
\epsscale{.60}
\plotone{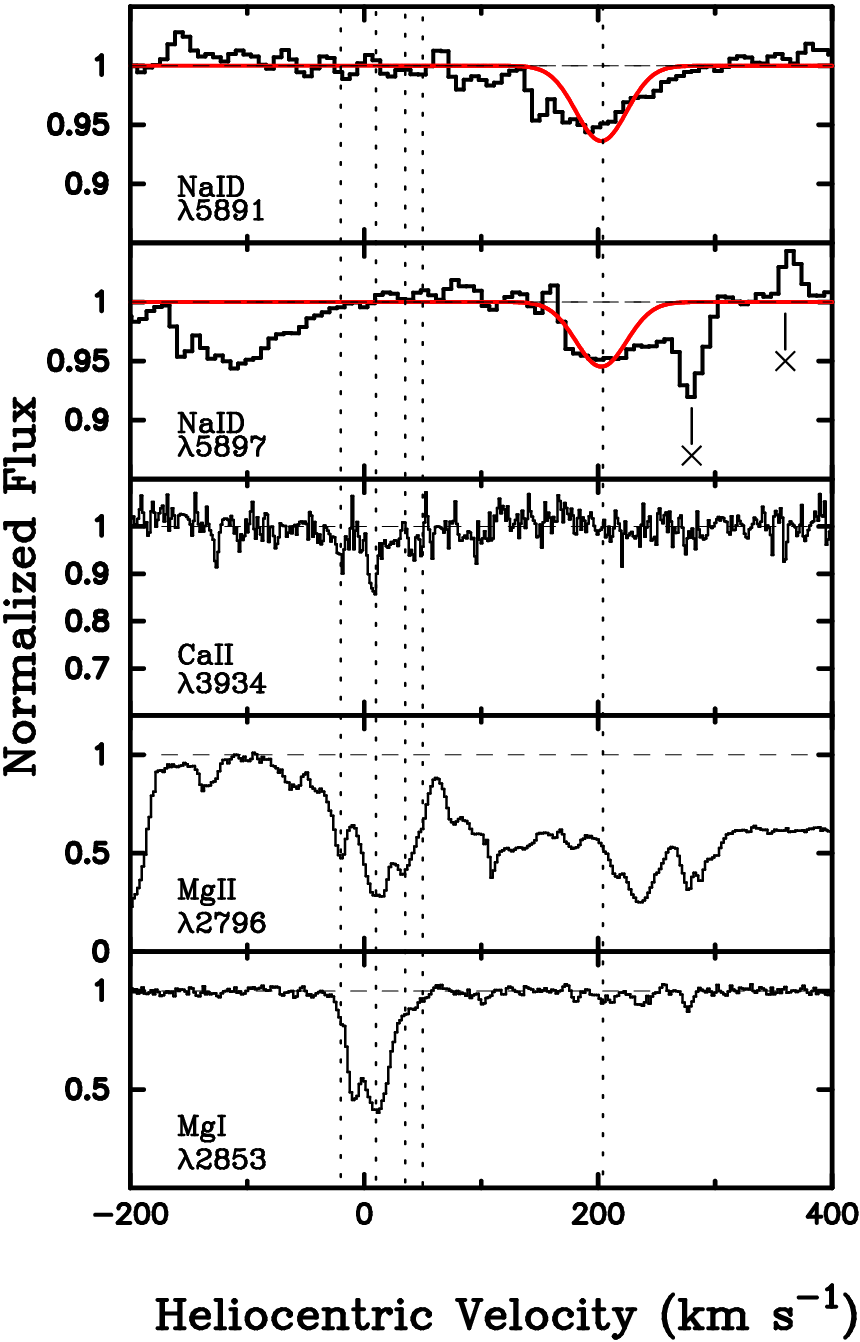}
 \caption{Velocity profile of metal lines form $\rm{z_{abs}=1.173}$
 system.  Notation are same as those in Figure 2. Horizontal axis shows
 the heliocentric velocity from z=1.1727. Bottom three panels show the
 velocity profiles of \caii\ $\lambda 3934$, \mgii\ $\lambda 2803$ and
 {\rm Mg}~{\sc i} $\lambda 2852$ absorption lines obtained with
 HIRES. Vertical dashed lines mark velocity components
 $\rm{(V=-20,0,+35,+50\ km\ s^{-1})}$ found by \citet{Pet00} and those
 $\rm{(V=+204\ km\ s^{-1})}$ newly found with \nai\ D data. 
 Spurious features labeled with $\times$ mark on the right side of
 $\lambda$5897 absorption are due to hot pixels. See the main text for
 detail. \label{fig4}}
\end{figure}

  The fitting results are summarized in Table 1. Figures
\ref{fig2},\ref{fig3},\ref{fig4} show the velocity profiles and fitting
results for the \nai\ D lines. Using atmospheric
absorption lines in the object and standard spectra, we confirmed that
the pixel shift along the dispersion direction was less than 1 pixel
throughout the observing time. Therefore the systematic uncertainty of
the velocity is less than $\rm{7.5\ km\ s^{-1}}$.  Although the
column densities and the Doppler widths could not be directly determined
because of insufficient instrumental spectral resolution, they can also
be constrained by the doublet ratio [i.e., equivalent width ratio
$W(5891)/W(5897)$].  As the doublet ratio approaches unity, uncertainty
in column density increases because the column density and Doppler width
become degenerate for a fixed equivalent width.

Because the physical distance ($\sim$ 1 kpc) between A and B at
$z\sim1$ is much larger than the typical 
spatial scale of \nai\ clouds in our Galaxy \citep[a few tens to several
thousands AU e.g.,][]{Pri01}, the images
A and B are most likely covered unequally.  This introduces an
additional systematic uncertainty in the inferred properties of the
\nai\ clouds.  An estimation of the systematic uncertainty depends
on the saturation level of the \nai\ absorption lines in the A+B
combined spectrum.  It can range from $\sim$0.3 dex for a moderately
saturated case to $\sim$ 1 dex for a heavily saturated case. The
estimated systematic uncertainty for each component is shown
in Table 1.  

Because the resolution of the optical spectrum is higher than our
near-infrared spectrum, the estimated redshift of each 
component by \citet{Pet00} should be more accurate than ours.
Hereafter, we use the velocities reported by \citet{Pet00} for our
discussion of each component, except for the newly identified
components from our spectrum.

\subsection{DLA at $\rm{z_{abs}=1.062}$}

In this system, we found two \nai\ components at $+53$ and $+108\
\rm{km\ s ^{-1}}$, which we identify as the components at $+55$ and
$+120\ \rm{km\ s^{-1}}$ by \citet{Pet00}.  While Petitjean et~al.\
found \nai\ component only at the +55 $\rm{km\ s^{-1}}$, we newly
identified \nai\ component at the +120\ $\rm{km\ s^{-1}}$.  The
estimated column densities are $\rm{\log N_{NaI}} = 12.9$ and $11.7$,
respectively.  These are consistent with Petitjean et~al., who
detected the former component with the similar column density, but not
the latter component to an upper-limit of $\rm{\log N_{NaI}}<12.8$.

\subsection{DLA at $\rm{z_{abs}=1.181}$}

In this system, we found five \nai\ components.  We identified the
three components at $-99$, $-75$, and $+48\ {\rm km\ s^{-1}}$ as the
components at $-105$, $-80$, and $+40\ \rm{km\ s^{-1}}$ reported by
\citet{Pet00}. While Petitjean et~al.\ found \nai\ component only at  the $-80\
\rm{km\ s^{-1}}$, we newly identified \nai\ components at the $-105$,
and $+40\ {\rm km\ s^{-1}}$. We identified further two new
\nai\ components at $+4$, and $+101\ \rm{km\ s^{-1}}$, which were not
identified in Petitjean et~al\ probably due to contamination by other
absorption lines.  The estimated column density of the $-80\ \rm{km \
s^{-1}}$ component is $\rm{log\ N_{NaI} =12.5}$, which is not
consistent with Petitjean et~al.  
However, our evaluation of the column density of $-80\ \rm{km\ s^{-1}}$
component is not so robust because \mgi\
$\lambda 2852$ absorption associated with an \mgii\ absorption system
at $\rm{z_{abs}}=3.502$ (Kondo et~al.\ in preparation) overlaps with
the \nai\ D $\lambda 5891$.  The estimated column density of the
$-105\ \rm{km \ s^{-1}}$ component is $\rm{log\ N_{NaI} =12.0}$, which
is consistent with the upper-limit of $\rm{log\ N_{NaI}<12.4}$ found
by Petitjean et~al.

\subsection{\mgii\ system at $\rm{z_{abs}=1.173}$}

In this system, we found a \nai\ component at $+204\ \rm{km\ s
^{-1}}$. \citet{Pet00} did not find any metal absorption lines at this
velocity. The detected \nai\ absorption line probably arises in a
cloud that was previously not in front of the quasar at this velocity.
We discuss the detail of this newly identified component in \S 4.2.
We could not detect \nai\ absorption lines associated with the four
components around $\rm{0\ km\ s^{-1}}$ reported by Petitjean et~al.,
probably because of their low \nai\ column density.

\section{Discussion}

\subsection{\hi\ Column Densities of Each Velocity Component}

Following \citet{Pet00}, we estimated $\rm{N_{HI}}$ for all 
components using the empirical relationship, $\rm{\log\
N_{HI}=0.688\,\log\, N_{NaI}+12.16\ (11.6<\log N_{NaI} < 13.4)}$
\citep{Sem93,Dil94,Bow95} established for the Galactic interstellar
absorption clouds.  In Table 2, we summarize our results for
$\rm{N_{HI}}$. While this empirical relationship does not hold for
$\rm{\log \ N_{NaI} <11}$ \citep{Welt94,Wak00,Wak01} nor for $\rm{\log\
N_{NaI}>12.9}$ \citep{Fer85}, the column densities of all the detected
components are within the appropriate range of this relationship. The estimated $\rm{N_{HI}}$ in various components, however, is subject
to systematic uncertainties related to the following issues:

\begin{enumerate}
\item Low metallicity

It is likely that the column density ratio $\rm{N_{NaI}/N_{HI}}$ is in
proportion to the metallicity. \citet{Vla93} found that the column
density ratio in the Large Magellanic Cloud (LMC) decreases in
proportion to the metallicity \cite[$\sim 0.3\
Z_{\odot}$,][]{Pei74,Pag77}. Because the metallicities of the
$\rm{z_{abs}}=1.062$ and $1.181$ DLAs are estimated to be in the
range of $0.3 - 1.0\, Z_{\odot}$ \citep{Pet00}, it is appropriate to
interpret the estimated $\rm{N_{HI}}$ in these systems as
lower-limits.

\item Contamination from a metal absorption line of other redshift

We could not evaluate the reliable column density of the $-80\ {\rm
km\ s^{-1}}$ component in the $\rm{z_{abs}}=1.181$ system because
\mgi\ $\lambda 2852$ absorption associated with a strong \mgii\ system
at $\rm{z_{abs}}=3.502$ (Kondo et al. in preparation) is blended with
the \nai\ D $\lambda\ 5891$ absorption. Assuming the correlation,
$\rm{\log\ N_{MgI}/N_{MgII}=-0.73\, \log\ N_{MgII}+7.6}$ \citep{Chu03}
and the same Doppler width $\rm{(b=13.7\ km\ s^{-1})}$ as the \mgii\
absorption lines, we estimated the equivalent width of the \mgi\
$\lambda 2852$ absorption.  The \hi\ column density of this component
was estimated after removing this contamination from \mgi, and this
introduced systematic uncertainty. Even after removing the
contamination, there still is another large uncertainty, because the
resultant doublet ratio of the \nai\ absorption lines is close to
unity.

\item Coverage of gravitationally-lensed images A and B

As stated above, it is likely that the \nai\ clouds do not equally cover the
gravitationally-lensed images A and B.  Thus, the estimated \nai\
column density  is uncertain as described in \S 3. The estimated $\rm
{N_{HI}}$ varies accordingly.

\end{enumerate}

While \citet{Pet00} suggested that there are 40 and 20 small clouds in
the $\rm{z_{abs}}=1.062$, and 1.181 systems, respectively, our results
show that there are at least 2 and 5 high column density clouds in each
system. Most interestingly, we find that each cloud in the
$\rm{z_{abs}}=1.181$ system has a \hi\ column density that is classified
as a DLA. This is the first indication of an absorber in which the {\it
individual\/} clouds are each of such high column density.  This is in
contrast to the findings of \citet{Chu03-dla}, who constrained the size
of DLA clouds be less than 25 pc in the lensed QSO 0957+561A,B and
inferred that only one of the clouds gave rise to the DLA \hi\ column
density.  This would suggest that the $\rm{z_{abs}}=1.181$ DLA, having
$\log\, N_{\rm{HI}} \sim21.2$, may be probing a very massive and large
underlying structure \citep[see, for example, ][]{Tur04}.

\subsection{Detection of a New DLAs at $\rm{z_{abs}=1.173}$:\ 
Proper Motion of a cloud of a few 100 AU?}

We detected a new component at $\rm{+204\ km\ s^{-1}}$ associated with
the $\rm{z_{abs}=1.173}$ \mgii\ absorption system.  Based upon scaling
from the high \nai\ column density, this system is probably a DLA
rather than a Lyman-limit \mgii\ system \citep{Pet00}. Therefore, this
system might contribute to the gravitational-lensing of the QSO,
together with the other three DLAs, at $\rm{z_{abs}}= 1.062,\ 1.181,\
and\ 2.974$, as suggested by Petitjean et~al.  Despite the large \nai\
column density, Petitjean et~al.\ did not identify any metal lines for
this velocity component in the optical Keck spectrum. Similarly, in a
subsequent search, we also could not identify any \mgii, \mgi,\ and \caii\
absorption lines for this component in the HIRES
spectrum\footnote{ftp://ftp.ast.cam.ac.uk/pub/papers/APM08279} (see
Figure \ref{fig4}).  

Because there is a time-span of 3.6 years between our
observation and the optical observation in the local-frame, or $\sim$1.7
years in the rest-frame $\rm{(z_{abs}=1.173)}$ considering the effect of
cosmological time dilation by a factor of $\rm{(1+z_{abs})^{-1}}$, it is
probable that the newly identified cloud appeared in the line of sight
toward the QSO within that time-span. If the cloud is moving in a
tangential direction in the sky with a velocity of $\rm{\sim 400\ km\
s^{-1}}$, the travel distance is $\sim 200$\ AU in $\sim$1.7 years in
the rest-frame. The assumed velocity is comparable to the combined
velocity of the motion of cloud in high-redshift galaxies
\citep[$\rm{\sim 50\ to \sim 260\ km\ s^{-1}}$,][]{erb03} and the proper
motion of field galaxies which should be smaller than the typical
velocity dispersion of the cluster of galaxies \citep[e.g.,
$\rm{between\sim 500\ and \sim 1000\ km\ s^{-1}}$,][]{Gal04}.  This sets
the upper-limit of the cloud size and it is consistent with the size of
the \nai\ clouds in the Galaxy, which ranges from a few to several
thousands AU \citep[e.g.,][]{Mey96,Mey99,Welt01,And01,Lau00,Lau03}.
Our result shows that we can catch the very small-scale
structures of the cold $\rm{(\sim\ 100\, K)}$ neutral gas component in
high-z QSO absorption systems with time variation of \nai\ absorption
features. The observable spatial scale with this technique is much
smaller than that with the observation of gravitationally-lensed QSOs
\citep[see discussion in][]{Ell04}.

\subsection{Dust Depletion and Chemical Uniformity in DLAs}

The column density ratio $\rm{N_{NaI}/N_{CaII}}$ has often been used
as a good indicator of the degree of dust depletion because sodium is hardly
depleted from gas onto dust grains while calcium can be heavily
depleted \citep{Sav96}.  In Table 3, we list the ratio
$\rm{N_{NaI}/N_{CaII}}$ of each component.  $\rm{N_{CaII}}$ for the
$\rm{+120\ km\ s^{-1}}$ component at $\rm{z_{abs}=1.062}$ and for the
$-105$ and $-80\ {\rm km\ s^{-1}}$ components at $\rm{z_{abs}=1.181}$
are taken from \citet{Pet00}, while those for the $+4$, $+40$, and
$+101\ {\rm km\ s^{-1}}$ components at $\rm{z_{abs}=1.181}$ were
evaluated\footnote{We evaluated \caii\ column density for those
systems using VPFIT.  We evaluated \caii\ column densities also for
the $+120\ {\rm km\ s^{-1}}$ component at $\rm{z_{abs}=1.062}$ and the
$-105$, and $-80\ {\rm km\ s^{-1}}$ components at $\rm{z_{abs}=1.181}$
and confirmed that our fitting results are consistent with those
estimated by \citet{Pet00}.} using archived optical spectra
\citep{Ell99a,Ell99b}.  

We estimated the 3$\sigma$ upper-limit of $\rm{N_{CaII}}$ for the
$+4$, and $+101\ {\rm km\ s^{-1}}$ components at $\rm{z_{abs}=1.181}$,
because no \caii\ absorption line was detected in the optical spectra.
In Figure \ref{fig4}, we show a $\rm{\log\, [N_{NaI}/N_{CaII}]}$ -- $\rm{\log\
N_{NaI}}$ plot for the DLAs as well as \nai\ clouds in the Galaxy
\citep{Val93} and in the LMC \citep{Vla93}.

We found that $\rm{\log\, [N_{NaI}/N_{CaII}]}$ in these DLAs is
systematically smaller than those in the Galaxy but similar to those
in the LMC \citep{Welt99}, suggesting that dust depletion is generally
smaller in a lower metallicity environment. Our results are consistent
with the results of \citet{Vla98}, who shown that dust-to-metal
ratios of DLAs are similar to that of the LMC but smaller than that of
the Galaxy, using the column density ratio of zinc to chromium.

It is also interesting to consider the uniformity in the
${\rm N_{NaI}/N_{CaII}}$ ratio across the $\sim 200$ km~s$^{-1}$ velocity
spread of the $\rm{z_{abs}=1.181}$ DLA.  As presented in Table 3, it
would appear that the ratios are consistent with being uniform with
component velocity, though there is large uncertainty in this
statement.  If so, this is consistent with the findings of
\citet{Pro03-dla} and \citet{Chu03-dla}, who reported uniform column
density ratios as a function of velocity in DLAs.  It is also found by
Churchill et~al.\ that the uniformity is spatial, at least on the
scale of 200 pc.  If the ${\rm N_{NaI}/N_{CaII}}$ ratio is more--or--less
constant across the profiles in the $\rm{z_{abs}=1.181}$ DLA, it would
suggest that the chemical, ionization, and dust depletion levels are
uniform across the clouds.

\subsection{The $\rm{z_{abs}=1.181}$ System:\ 
Cold Dense Gas Clouds near the Center of a Galaxy ?}

All five components of the $\rm{z_{abs}=1.181}$ system have
$\rm{\log\, [N_{NaI}/N_{CaII}]>0}$, which is a characteristic of cold
dense gas in the Galaxy \citep[known as the ``the Routly-Spitzer
effect'', e.g.,][]{Rou52,Val93,Sem94}.  Because there are five cold
dense and very high column density clouds for the $\rm{z_{abs}=1.181}$
system, we suggest that the line of sight may be passing near the
center of the host galaxy.  The relatively high metallicity of this
system \citep[$0.3-1.0\, Z_{\odot}$,][]{Pet00} compared to that
typical of z$\sim$1 DLAs \citep[$\sim 0.1 \, Z_{\odot}$,][]{Pro03} also
supports the above idea, because the metallicity of DLA absorption is
higher in the inner region galaxies because of a metallicity
gradient \citep{Che05}.  Therefore, the host galaxy of this system
might be the most significant contributor to the gravitational-lens
among four DLAs toward APM 08279+5255.  Note that some DLAs were
observed to have an impact parameter larger than 65 kpc \citep{Chu05}
and above argument should be taken with caution.

\acknowledgments

We are grateful to all of the IRCS and AO team members and the Subaru
Telescope observing staffs for their efforts, which made it possible to
obtain these data.  We are also grateful to Ellison et al. for making
their excellent data available in the Internet and to Jochen Liske for
kindly answering our questions about VPGUESS.  We thank the National
Astronomical Observatory of Japan for their financial support and
encouragement for the construction of the IRCS and AO. Y.M. is financially supported by the Japan Society for the Promotion of Science
(JSPS).

\clearpage

\clearpage

\clearpage

\clearpage

\clearpage

\begin{figure}
\epsscale{.80}
\plotone{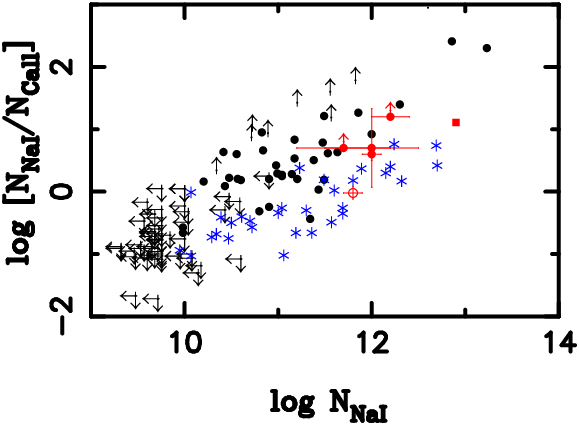}
\caption{The relation of $\rm{\log\, N_{NaI}\ and \log\,
 [N_{NaI}/N_{CaII}]}$.  Red points show our DLA data for
 $\rm{z_{abs}=1.181}$. A red circle shows our DLA data for
 $\rm{z_{abs}}=1.062$. Black and blue points show the data for our
 Galaxy \citep{Val93} and the LMC \citep{Vla93} , respectively. The
 uncertainty of those data is less than 20\%. Red filled square shows
 the data for the $\rm{V=-80\ km\ s^{-1}}$ velocity component at
 $\rm{z_{abs}=1.181}$, whose uncertainty is not shown because it could
 not be accurately estimated due to the accidental overlap of the
 $\rm{z_{abs}=3.502}$ \mgi\ absorption line (see main text).\label{fig5}}
\end{figure}

\clearpage

\begin{deluxetable}{crrrrcccc}
\tabletypesize{\footnotesize}
\tablewidth{0pt}
\tablecaption{The velocity components of the detected \nai\ D  absorption lines}

\tablehead{
\colhead{}&
\colhead{}& 
\multicolumn{3}{c}{Our results} &
\colhead{}& 
\multicolumn{3}{c}{\citet{Pet00}}  \\ 
\cline{3-5} \cline{7-9}

\colhead{system} &
\colhead{{$\rm{z_{abs}}$}} & 
\colhead{V\tablenotemark{a,b}}&
\colhead{$\rm{\log N_{NaI}}$} & 
\colhead{b} &
\colhead{}&
\colhead{V{\tablenotemark{a,c}}}& 
\colhead{$\rm{\log N_{NaI}}$} & 
\colhead{b}\\

\colhead{} &
\colhead{} & 
\colhead{($\rm{km\ s^{-1}}$)} & 
\colhead{}&
\colhead{($\rm{km\ s^{-1}}$)} &
\colhead{}& 
\colhead{($\rm{km\ s^{-1}}$)} &
\colhead{} &
\colhead{($\rm{km\ s^{-1}}$)}
}
\startdata
1.062& 1.06266$\pm 0.00001$ &  +53$\pm 2$&12.9\tablenotemark{d}$\pm
 1.6$&1.1$\pm 0.7$ & & +55 & 12.9&1.5\\   
1.062& 1.06305$\pm 0.00002$ & +108$\pm 3$&11.7\tablenotemark{e}$\pm
 0.1$&  7$\pm 12$&
 & +120& $<12.8$&1.5\\ \hline
1.173&  1.17418$\pm 0.00001$ &+204$\pm 1$& 12.3\tablenotemark{d}$\pm
 0.6$&1.0$\pm 0.4$ &
 &$\cdots$ &$\cdots$&$\cdots$\\ \hline
1.181& 1.17998$\pm 0.00006$&-99$\pm 8$& 12.0\tablenotemark{d}
 $\pm 0.5$ &  1.3$\pm 2$  & &-105&$<12.4$&2.5\\
1.181& 1.18015$\pm 0.00003$ &-75$\pm 4$& 12.5\tablenotemark{f}$\pm 0.1$&
 4$\pm 2$&  &-80&13.5&0.8\\
1.181& 1.18073$\pm 0.00005$ &+4$\pm 6$& 11.7\tablenotemark{f}$\pm 0.5$ &
 0.8$\pm 1$&  &$\cdots$&$\cdots$&$\cdots$\\
1.181&  1.18105$\pm 0.00003$ &+48$\pm 4$& 12.0\tablenotemark{e}$\pm 0.1$
 &  3.6$\pm 3$ & &+40&$\cdots$&$\cdots$\\
1.181&  1.18143$\pm 0.00002$ &+101$\pm 2$&
 12.2\tablenotemark{e}$\pm 0.2$ &  1.8$\pm 0.6$& &$\cdots$&$\cdots$&$\cdots$
\enddata
\tablenotetext{a}{The velocity zero point is set to z=1.06230, 1.17270,
 1.18070 for each system.}
\tablenotetext{b}{There is a systematic uncertainty
 which is less than $\rm{\sim7.5\ km\ s^{-1}}$. See main text for detail.}
\tablenotetext{c}{We estimated the velocity with an accuracy of
 $\rm{5\ km\ s^{-1}}$ from Figure 7,8 of \citet{Pet00} by eyes because
 \citet{Pet00} do not list the velocities.}
\tablenotetext{d}{There is a systematic uncertainty of about 1 dex
 because the gravitationally-lensed images are not resolved. See main
 text for detail.}
\tablenotetext{e}{There is an additional systematic uncertainty of
 about 0.3 dex because the gravitationally-lensed images are not
 resolved. See main text for detail.}
\tablenotetext{f}{There is a possible significant systematic
 uncertainty because of the accidental overlap of the
 $\rm{z_{abs}=3.502}$ \mgi\ absorption line. See main text for
 detail.}
\end{deluxetable}

\ProvidesFile{table2.tex}%
 [2003/12/12 5.2/AAS markup document class]%

\begin{deluxetable}{ccccccc}
\tablecaption{Estimated $\rm{N_{HI}}$ for the \nai\ D absorption lines}
\tablewidth{0pt}
\tablehead{\colhead{}& 
\colhead{}& 
\multicolumn{2}{c}{Our results} &
\colhead{}& 
\multicolumn{2}{c}{\citet{Pet00}}\\ \cline{3-4} \cline{6-7}

\colhead{system}& 
\colhead{V}& 
\colhead{$\rm{\log N_{NaI}}$}& 
\colhead{$\rm{\log N_{HI}}$}& 
\colhead{}&
\colhead{$\rm{\log N_{NaI}}$}& 
\colhead{$\rm{\log N_{HI}}$}}
\startdata
1.062&+55\tablenotemark{a}&12.9\tablenotemark{b}$\pm1.6$ &21.0\tablenotemark{b}$\pm1.1$& &$12.91\pm 0.04$&$21.0 	\pm0.0$  \\
1.062&+120\tablenotemark{a}&11.6\tablenotemark{c}$\pm 0.1$&20.2\tablenotemark{c}$\pm
 0.1$& &$<12.8$& $<21.0$ \\ \hline
1.173&+204&12.3\tablenotemark{b}$\pm0.6$&20.6\tablenotemark{b}$\pm 0.4$&
 &$\cdots$&$\cdots$ \\ \hline
1.181&-105\tablenotemark{a}&12.0\tablenotemark{b}$\pm0.5$&20.4\tablenotemark{b}$\pm0.3$& &$<12.40$&$<20.7$\\
1.181&-80\tablenotemark{a}&12.9\tablenotemark{d}&20.7\tablenotemark{d}& &$13.5\pm0.11$& $21.4\pm 0.1$\\
1.181&+4&11.7\tablenotemark{c}$\pm 0.5$&20.2\tablenotemark{c}$\pm 0.3$& &$\cdots$&$\cdots$\\
1.181&+40\tablenotemark{a}&12.0\tablenotemark{c}$\pm 0.1$&20.4\tablenotemark{c}$\pm0.1$& &$\cdots$&$\cdots$\\
1.181&+101&12.2\tablenotemark{c}$\pm0.2$&20.5\tablenotemark{c}$\pm0.1$& &$\cdots$&$\cdots$\\
\enddata
\tablenotetext{a}{Velocities from figures in \citet{Pet00}. See note b
 in Table 1 for detail.}
\tablenotetext{b}{There is an additional systematic uncertainty of
about 1 dex. See Table 1.}
\tablenotetext{c}{There is an additional systematic uncertainty of
about 0.3 dex. See Table 1.}
\tablenotetext{d}{There is a possible significant systematic
uncertainty. The value shown here is after removing of the
contamination from the \mgi\ absorption at $\rm{z_{abs}=3.502}$.}
\end{deluxetable}

\begin{deluxetable}{ccccc}
\tabletypesize{\small}
\tablecaption{Comparison of  $\rm{\log\, N_{NaI}}$ and $\rm{\log\, 
 [N_{NaI}/N_{CaII}]}$ for the \nai\ D absorption lines}
\tablewidth{0pt}
\tablehead{
\colhead{system}& 
\colhead{V}& 
\colhead{$\rm{\log\, N_{NaI}}$}&
\colhead{$\rm{\log\, N_{CaII}}$}& 
\colhead{$\rm{\log\, [N_{NaI}/N_{CaII}]}$}} 
\startdata
1.062&+50&12.9\tablenotemark{a}$\pm1.6$ & $\cdots$\tablenotemark{d}&$\cdots$\tablenotemark{d}\\ 
1.062&+110&11.7\tablenotemark{b}$\pm 0.1$&$11.73\pm 0.03$&$-0.02\pm 0.1$ \\ \hline
1.181&-100&12.0\tablenotemark{a}$\pm0.5$&$11.33\pm 0.39 $&$0.7\pm0.6$\\
1.181&-75&12.9\tablenotemark{c}&$11.79\pm 0.35$&1.1\\
1.181&+4&11.7\tablenotemark{b}$\pm 0.5$&$<11$&$>0.7$\\
1.181&+40&12.0\tablenotemark{b}$\pm 0.1$&$11.4\pm 0.2$&$0.6\pm0.1$\\
1.181&+101&12.2\tablenotemark{b}$\pm0.2$&$<11$&$>1.2$\\
\enddata
\tablenotetext{a}{There is an additional systematic uncertainty of
 about 1 dex. See Table 1.}
\tablenotetext{b}{There is an additional systematic uncertainty of
 about 0.3 dex. See Table 1.}
 \tablenotetext{c}{There is a possibly significant systematic
 uncertainty.  The value shown here is after removing the
 contamination from the \mgi\ absorption at $\rm{z_{abs}=3.502}$.}
 \tablenotetext{d}{No measurement was possible due to the blending
 effects. See the note of TABLE 2 in \citet{Pet00}.}
\end{deluxetable}

\end{document}